\documentclass[copyright]{eptcs}
\providecommand{\event}{DCFS 2010} 
\usepackage{breakurl}             

\usepackage{latexsym}
\usepackage[dvips]{graphics}
\usepackage{epsfig}
\usepackage{amssymb,amsmath}
\usepackage[dvips]{graphics}
\usepackage{epsfig}
\usepackage{gastex}
\usepackage[usenames]{color}

\def\bbbn{{\rm I\!N}} 

\def \prend{\vrule depth-1pt height7pt width6pt}
\def \proof{\bigbreak\noindent{\bf Proof.\ \ }}
\def \endpf{{\ \ \prend \medbreak}}

\def\bbbn{{\rm I\!N}} 

\usepackage{amssymb}
\usepackage{graphicx}

\newtheorem{theorem}{T\/heorem}[section]

\newtheorem{corollary}{Corollary}[section]

\newtheorem{example}{Example}[section]

\newtheorem{proposition}[theorem]{Proposition}

\newtheorem{open}{Open problem}
\newtheorem{conjecture}[theorem]{Conjecture}

\begin{document}

\title{Finite-State Complexity and the Size of Transducers}

\author{Cristian S. Calude\footnote{Research supported in part by FRDF Grant
of the UoA.}\\
\institute{Department of Computer Science\\
University of Auckland, New Zealand}
\email{c.calude@auckland.ac.nz}
\and
Kai Salomaa\footnote{Research supported in part by NSERC.}\\
\institute{School of Computing\\ Queen's University\\
Kingston, Ontario, Canada}
\email{ksalomaa@cs.queensu.ca}
\and
Tania K. Roblot\\
\institute{Department of Computer Science\\
University of Auckland, New Zealand}
\email{trob048@aucklanduni.ac.nz}
}

\def\titlerunning{Finite-State Complexity}

\def\authorrunning{C.S.~Calude, K.~Salomaa, T.K.~Roblot}

\maketitle

\begin{abstract}
Finite-state complexity is a variant of algorithmic
information theory obtained by replacing Turing machines
with finite transducers.  We consider the state-size of
transducers needed for minimal descriptions of arbitrary
strings and, as our main result, show that the state-size
hierarchy with respect to a standard encoding is
infinite. We consider also hierarchies yielded by
more general computable encodings.\\
Keywords: finite transducers, descriptional complexity,
state-size hierarchy, computability 
\end{abstract}

\section{Introduction}
\label{intro}

Algorithmic information theory  \cite{Greg,cris}
uses the {\em minimal size\/}  of a Turing machine
that outputs a string $x$ as a descriptional complexity
measure. The theory has produced many elegant and important
results;  however, a drawback is that all  variants
of descriptional complexity 
based on various types of universal Turing
machines are incomputable. 
Descriptional complexity defined by resource
bounded Turing machines has been considered in~\cite{BF},
and, at the other end of the spectrum, lie models
based on context-free grammars or finite automata.

Grammar-based complexity measures the size of the smallest
context-free grammar generating a single string.
This model has been investigated since the 70's, and
recently there has been renewed interest 
due to applications in text compression and 
connections with Lempel-Ziv codings,
see e.g.~\cite{LS,Ry2}; a general overview
of this area can be found in
\cite{Le}. 
The {\em automatic complexity\/} of a string~\cite{SW} is
defined as the smallest
number of states of a DFA (deterministic finite
automaton) that accepts $x$ and does not accept any
other string of length 
$|x|$. Note that a DFA  recognizing the singleton language
$\{ x \}$  always needs  $|x| + 1$ states, which is the reason
 the definition considers only strings of length $|x|$. 
{\em Automaticity\/}~\cite{AS,SB} is an analogous descriptional
complexity measure for languages.
The {\em finite-state dimension\/} is defined in terms
of computations of finite transducers on infinite
sequences, see e.g.~\cite{BHV,DLN}.

The  NFA (nondeterministic finite automaton)
based complexity 
of a string~\cite{CLL} can also be viewed as being defined in terms of finite
transducers that are called  ``NFAs with advice''
in~\cite{CLL}.
However, the model allows the advice
strings to be over an arbitrary alphabet with no penalty
in terms of complexity and, as observed in~\cite{CLL},
consequently the NFAs used for compression can always be
assumed to consist of only one state.

The finite-state complexity of a finite string $x$
was introduced recently~\cite{CSR} 
in terms of a finite transducer and a string $p$
such that the transducer on input $p$ outputs $x$.
Due to the non-existence of  universal 
transducers, the size of the transducer is
included as part of the descriptional complexity 
measure. We get different variants of the measure
by using different encodings of the transducers.

In our main result we establish that the measure results in
a rich hierarchy in the sense that
there is no a priori
upper bound for the number of states used by transducers
in  minimal descriptions of given strings.
The result applies to our
standard encoding, as well as to any other ``reasonable'' encoding
where a transducer is encoded by listing the productions
in some uniform way.

By the {\em state-size hierarchy\/} we refer to the
hierarchy of languages $L_{\leq m}$, $m \geq 1$,
consisting of strings where a minimal description uses
a transducer with at most $m$ states. We show that the
state-size hierarchy with respect to the standard encoding
is infinite; however, it remains an open question whether
the hierarchy is strict at every level.

In a more general setting,
the definition of finite-state complexity~\cite{CSR} allows
an arbitrary computable encoding of the transducers,
and properties of the state-size hierarchy
depend significantly on the particular encoding.
We establish that,  for
suitably chosen computable encodings,
every level of the state-size hierarchy can be strict.

\section{Preliminaries}
\label{prelim}

If $X$ is a finite set then $X^{*}$ is the set of all strings
(words) over $X$, with $\varepsilon$ denoting the empty string.
The length of  $x\in X^{*}$
is denoted by $|x|$.  We use $\subset$ to denote strict
set inclusion.

For all unexplained notions concerning transducers we
refer the reader to~\cite{Be,Yu}.
In the following, by a {\em transducer\/}
we mean a {\em (left) sequential
transducer\/} \cite{Be}, also called a {\em deterministic generalised
sequential machine\/}  \cite{Yu},
where both the input and output alphabets are $\{ 0, 1 \}$.
The set of all transducers is $\cal T_{\rm DGSM}$.

A transducer $T \in \cal T_{\rm DGSM}$ is denoted as
a triple $T = (Q, q_0, \Delta)$ where
$Q$
 is the finite set of states, $q_0 \in Q$ is the start state,
(all states of $Q$ are considered to be final),
and
\begin{equation}
\label{seqtrans}
\Delta : Q \times \{ 0, 1 \} \rightarrow Q \times \{ 0, 1 \}^*
\end{equation}
is the transition function. 
When a transducer is represented
as a figure, each transition $\Delta(q, i) = (p,w)$,
$q, p \in Q$, $i \in \{ 0, 1 \}$, $w \in \{ 0, 1 \}^*$, is
represented by an arrow with label $i / w$ from state $q$ to
state $p$, and $i$ (respectively, $w$) is called the input (respectively,
output) label of the transition. By the (state) {\em size\/} of $T$,
${\rm size}(T)$, we mean number   of states in the set $Q$.

The function $\{0,1\}^{*} \rightarrow \{0,1\}^{*}$ computed by
the transducer  $T$ is, by slight abuse of
notation, also denoted by $T$ and
defined by
$T(\varepsilon)= \varepsilon$,
$T(xa) = T(x) \cdot \pi_2(\Delta(\hat{\delta}(q_{0},x),a))$,
for
$x \in \{0,1\}^{*}$,
$a \in \{0,1\}$. Here $\pi_i$, $i = 1, 2$, are the
two projections on $Q \times \{ 0, 1 \}^*$,
and $\hat{\delta} :
Q \times \{ 0, 1 \}^*
\rightarrow  Q$ is defined by  $\hat{\delta} (q, \varepsilon)=q,
\hat{\delta} (q,xa) =
\pi_1(\Delta(\hat{\delta} (q,x),a))$, $q\in Q$, $x \in \{0,1\}^{*}$
$a \in \{ 0, 1 \}$.

By a  {\em computable encoding\/} of all transducers 
we mean a pair
$S = (D_S, f_S)$ where  $D_S \subseteq \{ 0, 1 \}^*$ is
a decidable set and $f_S: D_S \rightarrow {\cal T}_{DGSM}$
is a computable bijective mapping  that associates
a transducer
$T^S_\sigma$
to each $\sigma \in D_S$.\footnote{In
a more general setting the mapping $f_S$ may not be injective
(for example, if we want to define $D_S$ as a regular set~\cite{CSR}),
however, in the following we restrict consideration to bijective
encodings in order to avoid unnecessary complications with
the notation associated with our state-size hierarchy.}

We say that $S$ is a {\em polynomial-time
 (computable) encoding\/} if  $D_S \in P$ 
and for a given $\sigma \in D_S$ we can compute 
 the transducer $T^S_\sigma \in {\cal T}_{DGSM}$
in polynomial time.
We identify a transducer $T \in {\cal T}_{DGSM}$  with
its transition function~(\ref{seqtrans}), and the
set of state names
is always $\{ 1, \ldots, |Q| \}$ where $1$ is the start state.
By computing the transducer $T^S_\sigma$ we mean an
algorithm that (in polynomial time) outputs
the list of transitions (corresponding to~(\ref{seqtrans}),
with state names  written in binary) of
$T^S_\sigma$.

Next we define a fixed natural  encoding $S_0$ of transducers
that we call the {\em standard encoding}. For our main 
result we need some fixed encoding of the transducers where the
\label{statesize}
length of the encoding relates in a ``reasonable way'' to the
lengths  of the transition outputs.
We encode a transducer as a binary string by listing for each state
$q$ and input symbol $i \in \{ 0, 1 \}$ the output and target
state corresponding to the pair $(q, i)$, that is, $\Delta(q, i)$.
Thus, the encoding of a transducer is a list of (encodings
of) states and output strings. For succinctness,
in the list
we omit (that is,
replace by $\varepsilon$) the states that correspond to self-loops.

By bin($i$) we denote the
binary representation of $i \geq 1$. Note that for all $i \geq 1$,
${\rm bin}(i)$ always  begins with
a $1$.
For $v = v_1 \cdots v_m$,
$v_i \in \{ 0, 1 \}$, $i = 1, \ldots, m$, 
we use the following functions producing self-delimiting versions of
their inputs
(see \cite{cris}):
$v^{\dagger}= v_10v_20\cdots
v_{m-1}0v_m 1$ and $v^{\diamond}=
\overline{(1v)^{\dagger}}$, where
$\overline{\phantom{x}}$ is the negation morphism given
by  $\overline{0}=1, \overline{1}=0$. It is seen that
$|v^{\dagger}|=2|v|,$ and
$|v^{\diamond}|=2|v|+2$.

We define the set $D_{S_0}$  to consist of
all strings of the form
\begin{equation}
\label{latta3}
\sigma=
{\rm bin}(i_1)^{\ddagger} \cdot
v_1^{\diamond} \cdot
{\rm bin}(i_2)^{\ddagger} \cdot v_2^{\diamond}
\cdots {\rm bin}(i_{2n})^{\ddagger} \cdot v_{2n}^{\diamond},
\end{equation}
where $1 \leq i_t \leq n$, $v_t \in \{ 0, 1 \}^*$,
$t = 1, \ldots, 2n$,
and
$$
{\rm bin}(i_t)^{\ddagger} = \left \{
\begin{array}{l}
{\rm bin}(i_t)^{\dagger} \mbox{ if }  i_t \neq 
\lceil \frac{t}{2} \rceil, \\
\varepsilon \mbox{ if }  i_t  =
\lceil \frac{t}{2} \rceil. 
\end{array} \right., \;\;
1 \leq t \leq 2n.
$$

A string $\sigma$ as in~(\ref{latta3}) encodes
the transducer $T^{S_0}_\sigma = (\{1, \ldots, n \},  1, \Delta)$,
where
$\Delta(j, k) =
({i_{2j-1+k}}, v_{2j-1+k})$,
$j =  1, \ldots, n$, $k\in \{0,1\}$.
Note that in~(\ref{latta3}),
${\rm bin}(i_t)^{\ddagger}=\varepsilon$
if the corresponding transition of $\Delta$ is a self-loop.

Now we define the
standard encoding $S_0$ as the pair $(D_{S_0},
f_{S_0})$ where $f_{S_0}$ associates
to each $\sigma \in S_0$ the transducer 
$T^{S_0}_\sigma$ as described above.
It can be verified  that for each
$T \in {\cal T}_{DGSM}$ there exists a unique
$\sigma \in D_{S_0}$ such that $T = T^{S_0}_\sigma$,
that is, $T$ and $T^{S_0}_\sigma$ have the same
transition function.
The details of verifying that $T^{S_0}_{\sigma_1} \neq
T^{S_0}_{\sigma_2}$ when $\sigma_1 \neq \sigma_2$ can
be found in~\cite{CSR}. For  $T \in {\cal T}_{DGSM}$, 
the {\em standard encoding of\/} $T$ is the unique
$\sigma \in D_{S_0}$ such that $T = T^{S_0}_\sigma$.
The standard encoding  $S_0$ is a
polynomial-time encoding.


Note that using a modification of the above definitions
it is possible
to guarantee that the set of all legal encodings of transducers
is regular~\cite{CSR} -- this is useful e.g.,
for  showing that the
non-existence of a universal transducer is not caused simply
by the fact that a finite transducer cannot recognize legal
encodings of transducers. More details about
computable encodings can be found in~\cite{CSR}, including
 binary encodings that are more efficient
than the standard encoding.

\section{Finite-state complexity}
\label{fs-compl}

In the general form, the
transducer based finite-state complexity with respect
to a computable encoding $S$ of transducers in ${\cal T}_{DGSM}$
is defined as follows~\cite{CSR}.

We say that a pair $(T_{\sigma}^{S}, p)$, $\sigma \in D_S$, 
$p \in \{0,1\}^*$, 
defines
the string $x \in \{ 0, 1 \}^*$ provided that
$T_{\sigma}^{S}(p)=x$; the pair $(T_{\sigma}^{S}, p)$ is called a 
{\it description} of $x$.
As  the pair $(T_{\sigma}^{S}, p)$ is uniquely represented by 
the pair $(\sigma, p)$ we
define the {\em  size\/} 
of the description $(T_{\sigma}^{S}, p)$   by
$$||(T_{\sigma}^{S}, p)||_{S} = |\sigma| + |p|.$$
We define the {\em finite-state complexity} of a string 
$x \in \{ 0, 1 \}^*$
 with respect
to encoding $S$   by the formula:
$$
C_{S}(x) =    \inf_{\sigma \in D_S, \; p \in \{ 0, 1
\}^* } \Big\lbrace  \mid \sigma \mid + \mid p \mid \; : 
T_\sigma^{S} (p) = x \Big\rbrace.
$$

We will be interested in the state-size,
that is, the number of states of transducers
used for minimal encodings of arbitrary  strings.
For $m \geq 1$ we define the language $L^{S}_{\leq m}$ to consist of
strings $x$ that have a minimal description using a transducer with
at most $m$ states. Formally, we write
\begin{eqnarray*}
L^{S}_{\leq m} & =  \{ & x \in \{ 0, 1 \}^* \; \mid \;
(\exists \sigma \in D_S, p \in \{ 0, 1 \}^*) \;
T_\sigma^S(p) = x,\\
& &   |\sigma| + |p| = C_S(x),
{\rm size}(T_\sigma^S) \leq m \}.
\end{eqnarray*}
By setting $L^{S}_{ \leq 0} = \emptyset$,
the set of strings $x$ for which the
smallest number of states  of a transducer
in a minimal description of $x$ 
is  $m$ can 
then be denoted as
$$
L^{S}_{= m}  = L^{S}_{\leq m} - L^{S}_{\leq m-1}, \;\;\; m \geq 1.
$$
Also, we let $L^S_{\exists_{\rm min} m}$ denote
the set of strings $x$ that have a minimal description
in terms of a transducer with exactly $m$ states. 
Note that $L^{S}_{= m} \subseteq L^S_{\exists_{\rm min} m}$, but
the converse inclusion need not hold, in general, because
strings in $L^S_{\exists_{\rm min} m}$ may have other minimal
descriptions with fewer than $m$ states.

In the following,
when dealing with the standard
encoding $S_0$ (introduced in Section~\ref{prelim}) 
we write, for short, $T_\sigma$, $||(T,p)||$,
$C$ and $L_{\leq m}$,
$L_{= m}$, $L_{\exists_{\rm min} m}$,
$m \geq 1$,  instead
of $T^{S_0}_\sigma$, $||(T,p)||_{S_0}$,
$C_{S_0}$ and $L^{S_0}_{ \leq m}$,
$L^{S_0}_{ = m}$, $L^{S_0}_{\exists_{\rm min} m}$, respectively.
The main result in section~\ref{statessize} is proved using
the standard encoding; however, it   could easily be
modified  for any 
``naturally defined'' encoding of transducers,  
where each transducer is described by listing the states and transitions
in a uniform way. 
For example, the more efficient encoding considered in~\cite{CSR}
clearly satisfies this property. On the other hand, 
when dealing with arbitrarily defined computable encodings $S$,
the languages $L^{S}_{ \leq m}$, $m \geq 1$, obviously can have
very different properties. In  section~\ref{sec-viisi}
we will  consider properties of  more
general computable encodings.

The finite-state complexity with respect
to an arbitrary computable encoding $S$ is computable~\cite{CSR}
because for given $x$, $|\sigma_1| + |x|$ gives an upper bound
for $C_S(x)$ where $\sigma_1 \in S$ is an encoding of
the one-state identity transducer.
An encoding of the identity transducer
can be found from an enumeration
of strings in $S$, and after this we can simply try all transducer
encodings and input strings up to length $|\sigma_1| + |x|$. Hence
``inf'' can be replaced by ``min'' in the definition of $C_{S}$.

\begin{proposition}
\label{prop61}
For any computable encoding
$S$, the languages $L^{S}_{ \leq m}$, $m \geq 1$, are decidable.
\end{proposition}

We conclude this section with an example concerning the finite-state
complexity with respect to the standard encoding. 

\if01
\begin{figure}[ht]
\hspace*{\fill}
\epsfxsize=15cm
\epsfbox{hat3}
\hspace*{\fill}
\caption{
\label{ratta}
The transducer $T_1$ for Example~\ref{latta}.
In the transitions $\varepsilon$ denotes the empty string.
}
\end{figure}
\fi

\begin{example}
\label{latta}
{\normalfont 
Define the sequence of strings 
$$w_m = 1010^210^31 \cdot \ldots \cdot 0^{m-1}1 0^m 1, \;\; m \geq 1.$$
Using the transducer $T_1$ of Figure~\ref{ratta} we produce 
an encoding of $w_{99}$. Note that $|w_{99}| = 5050$.

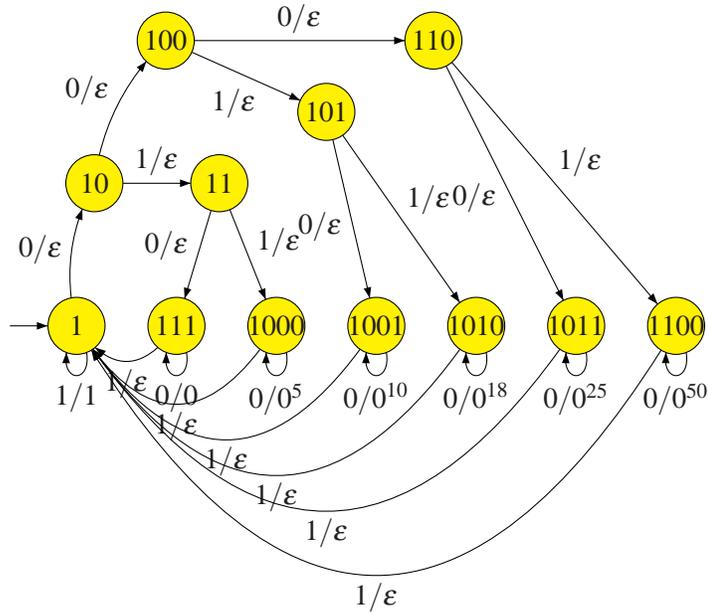
\begin{figure}[hbt]
\vspace{-5mm}
 \begin{center}
 \mbox{
   \unitlength=2.7pt
  \begin{picture}(71,58)(0,-9) 
 \gasset{Nw=8,Nh=8,Nmr=8,curvedepth=0}
   \thinlines
   \node[Nmarks=i,iangle=180,fillcolor=Yellow](A1)(0,0){$1$}
   \node[Nmarks=n,iangle=180,fillcolor=Yellow](A2)(14,0){$111$}
   \node[Nmarks=n,iangle=180,fillcolor=Yellow](A3)(28,0){$1000$}
   \node[Nmarks=n,iangle=180,fillcolor=Yellow](A4)(42,0){$1001$}
   \node[Nmarks=n,iangle=180,fillcolor=Yellow](A5)(56,0){$1010$}
   \node[Nmarks=n,iangle=180,fillcolor=Yellow](A6)(70,0){$1011$}
   \node[Nmarks=n,iangle=180,fillcolor=Yellow](A7)(84,0){$1100$}
   \node[Nmarks=n,iangle=180,fillcolor=Yellow](B1)(2.5,20){$10$}
   \node[Nmarks=n,iangle=180,fillcolor=Yellow](B2)(20,20){$11$}
   \node[Nmarks=n,iangle=180,fillcolor=Yellow](B3)(12.5,40){$100$}
   \node[Nmarks=n,iangle=180,fillcolor=Yellow](B4)(35,30){$101$}
   \node[Nmarks=n,iangle=180,fillcolor=Yellow](B5)(50,40){$110$}
   \gasset{loopdiam=3}
   \drawloop[loopangle=-90](A1){$1/1$}
   \drawloop[loopangle=-90](A2){$0/0$}
   \drawloop[loopangle=-90](A3){$0/0^5$}
   \drawloop[loopangle=-90](A4){$0/0^{10}$}
   \drawloop[loopangle=-90](A5){$0/0^{18}$}
   \drawloop[loopangle=-90](A6){$0/0^{25}$}
   \drawloop[loopangle=-90](A7){$0/0^{50}$}
   \drawedge(B1,B2){$1/\varepsilon$}
   \drawedge[ELside=r](B2,A2){$0/\varepsilon$}
   \drawedge(B2,A3){$1/\varepsilon$}
   \drawedge[ELside=r](B3,B4){$1/\varepsilon$}
   \drawedge(B3,B5){$0/\varepsilon$}
   \drawedge[ELside=r](B4,A4){$0/\varepsilon$}
   \drawedge(B4,A5){$1/\varepsilon$}
   \drawedge[ELside=r](B5,A6){$0/\varepsilon$}
   \drawedge(B5,A7){$1/\varepsilon$}
   \gasset{curvedepth=2}
   \drawedge(A1,B1){$0/\varepsilon$}
   \drawedge(B1,B3){$0/\varepsilon$}
   \gasset{curvedepth=5}
   \drawedge(A2,A1){$1/\varepsilon$}
   \gasset{curvedepth=11}
   \drawedge(A3,A1){$1/\varepsilon$}
   \gasset{curvedepth=16}
   \drawedge(A4,A1){$1/\varepsilon$}
   \gasset{curvedepth=21}
   \drawedge(A5,A1){$1/\varepsilon$}
   \gasset{curvedepth=26}
   \drawedge(A6,A1){$1/\varepsilon$}
   \gasset{curvedepth=35}
   \drawedge(A7,A1){$1/\varepsilon$}
   \end{picture}
    }
 \end{center}
 \vspace*{2.4cm}
\caption{The transducer $T_1$ for Example~\ref{latta}.
\vspace{0mm}
}
\label{ratta}
\end{figure}


With the encodings of the states
indicated in  Figure~1, $T_1$ is encoded by a string $\sigma_1
\in S_0$ of length 352. 
Each number $0 \leq i \leq 99$ can be represented as a
sum of, on average, 3.18 numbers from the multi-set
$\{ 1, 5, 10, 18, 25, 50 \}$~\cite{Sh}. Thus, when we represent
$w_{99}$ in the form $T_1(p_{99})$, we need on average at most
$6 \cdot 3.18$ symbols in $p_{99}$ to output each substring
$0^i$, $0 \leq i \leq 99$. (This is only a very rough estimate
since it assumes that for each element in the sum representing $i$
we need to make a cycle of length six through the start state, and
this is of course not true when the sum representing $i$ has some
element occurring more than once.) Additionally we need to produce
the 100 symbols ``1'', which means that the length of $p_{99}$ can
be chosen to be at most 2008.
Our estimate gives that
$$||(T_{\sigma_1}, p_{99})|| =  |\sigma_1| + |p_{99}| = 2360,$$
which is  a very rough upper bound for $C(w_{99})$.}
\end{example}

The above estimation could 
be improved using more detailed information
from the computation of the average from~\cite{Sh}. Furthermore, 
\cite{Sh} gives other systems of six numbers that,
on average, would give a more efficient way to represent numbers from 0
to 99 as the sum of the least number
of summands.\footnote{In \cite{Sh} it is established
that 18 is the optimal value to add to an existing
system of $\{ 1, 5, 10,  25, 50 \}$.} These types of constructions
can be seen to hint that computing the value of finite-state complexity
may have connections to the so-called postage stamp problems considered
in number theory,
with some  variants known
to be computationally
hard~\cite{Gu,Sh2}. It remains open whether computing the function
$C$ (corresponding to the standard encoding) is NP-hard, 
or more generally, whether for
some polynomial-time encoding $S$, computing
$C_S$ is NP-hard~\cite{CSR}.  

\section{State-size hierarchy }
\label{statessize}

We establish that finite-state complexity is a rich complexity measure
with respect to the number of states of the transducers, in the
sense that there is no a priori upper bound for the number of states
used for minimal descriptions of arbitrary strings.
This is in contrast to algorithmic information theory,
where the number of states of a
universal Turing machine can be fixed.

For the hierarchy result we use the standard encoding $S_0$.
The particular choice of the 
encoding  is not important and the proof could be easily modified
for any encoding that is based on listing the transitions of
a transducer in a uniform way.
However, as we will see later, arbitrary computable encodings can yield
hierarchies with very different properties.

\begin{theorem}
\label{se-ketta}
For any $n \in \bbbn$ there exists a string
$x_n$ such that $x_n \not\in L_{\leq n}$.
\end{theorem}
\proof
Consider an arbitrary but fixed $n \in \bbbn$. 
We define $2n+1$ strings of length $2n+3$,
$$
u_i = 10^i1^{2n+2-i}, \;\; i = 1,\ldots,2n+1.
$$
For $m \geq 1$, we define
$$
x_n(m) = u_1^{m^2} u_2^{m^2} \cdots u_{2n+1}^{m^2}.
$$

Let $(T_\sigma, p)$ be an arbitrary encoding of $x_n(m)$ where
${\rm size}(T_\sigma) \leq n$.
We show that by choosing $m$ to be sufficiently large as a function of
$n$, we have
\begin{equation}
\label{oatta}
||(T_\sigma, p)|| > \frac{m^2}{2}\raisebox{0.5ex}{.}
\end{equation}

The set of transitions of $T_\sigma$ can be written
as a disjoint union $\theta_1 \cup \theta_2 \cup \theta_3$,
where
\begin{itemize}
\item $\theta_1$ consists of the transitions where the output
contains a unique $u_i$, $1 \leq i \leq 2n+1$, as a 
substring,\footnote{By a substring we mean a ``continuous
substring''.}
that is, for any $j \neq i$, $u_j$ is not a substring of the output;
\item $\theta_2$ consists of the transitions where for distinct
$1 \leq i < j \leq 2n+1$, the output contains both $u_i$ and 
$u_j$ as a substring;
\item $\theta_3$ consists of transitions where the output
does not contain any of the $u_i$'s as a substring,
$i = 1, \ldots, 2n+1$.
\end{itemize}

Note that if a transition $\alpha \in \theta_3$ is used
in the computation $T_{\sigma}(p)$, the output produced by
$\alpha$ cannot completely overlap any of the occurrences
of $u_i$'s, $i = 1,\ldots,2n+1$.
Hence 
\begin{equation}
\label{oatta2}
\mbox{a transition of $\theta_3$  used  
by $T_{\sigma}$ on $p$ has output length at most $4n+4$.}
\end{equation}

Since $T_{\sigma}$ has at most $n$ states, and consequently at most $2n$
transitions, it follows by the
pigeon-hole principle that there exists $1 \leq k \leq 2n+1$ such that
$u_k$ is not a substring of any transition of $\theta_1$.
We consider how the computation
of $T_{\sigma}$ on $p$ outputs the substring $u_k^{m^2}$ of
$x_n(m)$. Let $z_1$, \ldots, $z_r$ be the minimal sequence of
outputs that ``covers'' $u_k^{m^2}$. That is, $z_1$ 
(respectively, $z_r$) is the output of a transition that
overlaps with a prefix (respectively, a suffix) of $u_k^{m^2}$
and $u_k^{m^2}$ is a substring of $z_1 \cdots z_r$. 

Define 
$$\Xi_i = \{ 1 \leq j \leq r \mid
z_j \mbox{ is output by a transition of } \theta_i \}, \;\; i =1,2,3.
$$
By the choice of $k$ we know that $\Xi_1 = \emptyset$.
For $j \in \Xi_2$, we know that the transition outputting
$z_j$ can be applied only once in the computation of
$T_{\sigma}$ on $p$ because for $i < j$ all occurrences
of  $u_i$ as substrings of $x_n(m)$ occur before all occurrences of $u_j$. 
Thus, for $j \in \Xi_2$, the use of this transition contributes
at least $2 \cdot |z_j|$ to the length of the encoding
$||(T_{\sigma},p)||$.

Finally, by~(\ref{oatta2}),
for any $j \in \Xi_3$ we have $|z_j| \leq 4n+4 < 2|u_k|$.
Such transitions may naturally be applied multiple times, however,
the use of each transition outputting $z_j$, $j \in \Xi_3$, contributes
at least one symbol to $p$. 

Thus, we get the following estimate:
$$
||(T_{\sigma}, p)|| \geq \sum_{j \in \Xi_2} 2 \cdot |z_j|
+ |\Xi_3| > \frac{|u_k^{m^2}|}{2|u_k|} = \frac{m^2}{2} \raisebox{0.5ex}{.}
$$

To complete the proof it is sufficient to
show that, with a suitable choice of $m$, 
$C(x_n(m)) < \frac{m^2}{2}$.
The string $x_n(m)$ can be represented by the pair
$(T_1, p_1)$ where $T_1$ is the $2n$-state transducer
from Figure~\ref{museol} and $p_1 = (0^m1)^{2n-1}0^m1^m$.

\if01
\begin{figure}[ht]
\hspace*{\fill}
\epsfxsize=14cm
\epsfbox{tat1}
\hspace*{\fill}
\caption{
\label{museol}
The transducer $T_0$ from the proof of Theorem~\ref{manystates}.
In the figure an output label $e$ refers to the empty
string.}
\end{figure}
\fi
\vspace*{1.5cm}

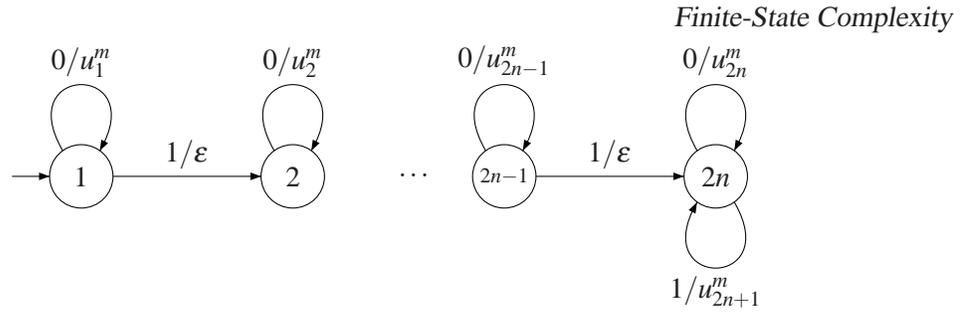
\begin{figure}[ht]
 \begin{center}
 \mbox{
     \unitlength=4pt
  \begin{picture}(51, 17)(0,-9)
 \gasset{Nw=6,Nh=6,Nmr=3,curvedepth=0}
   \thinlines
   \node[Nmarks=i,iangle=180](A1)(0,0){$1$}
   \node(A2)(20,0){$2$}
   \node(A3)(40,0){$\substack{2n-1}$}
   \node(A4)(60,0){$2n$}
   \drawloop[loopangle=90](A1){$0/u^m_1$}
   \drawedge(A1,A2){$1/\varepsilon$}
   \drawloop[loopangle=90](A2){$0/u^m_2$}
   \put(30,0){$\ldots$}
   \drawloop[loopangle=90](A3){$0/u^m_{2n-1}$}
   \drawedge(A3,A4){$1/\varepsilon$}
   \drawloop[loopangle=90](A4){$0/u^m_{2n}$}
   \drawloop[loopangle=-90](A4){$1/u^{m}_{2n+1}$}
   \end{picture}
   }
 \end{center}
 \medskip

\caption{
\label{museol}
The transducer $T_1$ from the proof of Theorem~\ref{se-ketta}.}
\end{figure}

Each state of $T_1$ can be encoded by a string
of length at most $\lceil \log_2 (2n) \rceil$,  so 
(recalling that in the standard encoding
each transition output $v$ contributes $|v^\diamond| = 2|v|+2$ to
the length of the encoding and
each binary encoding $u$ of a state name that is the
target of a transition that is not a self-loop contributes
$2|u|$ to the length of the encoding) we
get the following upper bound 
for the length of
a string $\sigma_1 \in S_0$ encoding $T_1$:
$$
|\sigma_1| \leq (8n^2 + 16n + 8)m + 
(4n-2)(\lceil \log_2(2n) \rceil + 1).
$$
Noting that $|p_1| = (2n+1)m + 2n-1$ we observe that
\begin{equation}
\label{hatta1}
C(x_n(m)) \leq  ||(T_{\sigma_1},p_1)|| = 
|\sigma_1| + |p_1|    < \frac{m^2}{2} \raisebox{0.5ex}{,}
\end{equation}
for example, if we choose $m = 16 n^2 + 36 n + 19$.
 This completes the proof.
\endpf

As a corollary we obtain that the sets of strings
with minimal descriptions using
a transducer   with  at most $m$ states, $m \geq 1$, form an infinite
hierarchy. 

\begin{corollary}
\label{se-ketta2}
For any $n \geq 1$, there exists effectively $k_n \geq 1$ such that
$L_{ \leq n} \subset L_{ \leq n + k_n}$.\footnote{Note that here
``$\subset$'' stands for
strict inclusion.}
\end{corollary}

We do not know whether all levels of the state-size hierarchy
with respect to the standard encoding are strict.
Note that the proof of Theorem~\ref{se-ketta} constructs
strings $x_n(m)$ 
that have a smaller description
using a transducer with $2n$ states than any description
using a transducer with $n$ states. 
We believe that (with $m$ chosen as in the proof of Theorem~\ref{se-ketta})
the minimal description of $x_n(m)$, in fact, has $2n$ states,
but do not have a complete proof for this claim.
The claim would imply
that $L_{\leq n}$ is strictly included in $L_{\leq 2n}$, $n \geq 1$.
In any case, the construction used in the proof of
Theorem~\ref{se-ketta} gives an effective upper bound for the
size of $k_n$ such that $L_{ \leq n} \subset L_{ \leq n + k_n}$,
because the estimation~(\ref{hatta1}) (with the particular
choice for $m$) implies also an
upper bound for the number of states
of a transducer used in a minimal description of $x_n(m)$.

The standard encoding is monotonic in the sense that  adding
states to a transducer or increasing the lengths of the outputs,
always increases the length of an encoding. This leads us to believe
that for any $n$ there exist strings where the minimal transducer has
exactly $n$ states, that is, 
for any $n \geq 1$, $L_{= n } \neq \emptyset$.

\begin{conjecture}
\label{se-ketta3}
$L_{\leq n} \subset L_{\leq n+1}$, for all $n \geq 1$.
\end{conjecture}

By Proposition~\ref{prop61} we know that the languages
$L_{\leq n}$ are decidable. 
Thus, for $n \geq 1$ such  that $L_{=n} \neq \emptyset$, 
in principle, it would be possible to compute
the  length $\ell_n$ of shortest words in $L_{=n}$.
However, we do not know how
$\ell_n$ behaves as a function of $n$. Using a brute-force
search we have established~\cite{CSR} that all strings of length at most
23 have a minimal description  using a single
state transducer.

\begin{open}
\label{pituus}
What is the asymptotic behavior of the length of the shortest words
in $L_{=n}$ 
as a function
of $n$?
\end{open}

Also, we do not know whether there exists $x \in \{ 0, 1 \}^*$
that has two minimal descriptions 
(in the standard encoding) where the respective transducers
have different numbers of states. This amounts to the following:

\begin{open}
\label{se-ketta4}
Does there exist $n \geq 1$ such that
$L_{=n} \neq L_{\exists_{\rm min} n}$?
\end{open}

\section{General computable encodings}
\label{sec-viisi}

While the proof of Theorem~\ref{se-ketta} can be easily modified
for any encoding that, roughly speaking, is based on listing
the transitions of a transducer, the proof breaks down if we consider
arbitrary computable encodings $S$.
Note that the number of transducers with $n$ states is infinite
and, for arbitrary computable $S$, it does not seem easy, 
analogously as in the proof of Theorem~\ref{se-ketta}, to get
upper and lower bounds for $C_S(x_n(m))$ for suitably chosen
strings $x_n(m)$.
We do not know whether there exist computable encodings
for which the state-size hierarchy collapses to a finite level.

\begin{open}
\label{se-tatta}
Does there exist $n \geq 1$ and a computable encoding $S_n$ such that
that, for all $k \geq 1$, $L^{S_n}_{\leq n} = L^{S_n}_{\leq n + k}$?
\end{open}

On the other hand, it is possible to construct particular
encodings 
for which every level of the state-size hierarchy 
is strict.

\begin{theorem}
\label{se-tatta2}
There exists a computable encoding
$S_{1}$ such that
$$
L^{S_{1}}_{ \leq n-1} \subset L^{S_{1}}_{ \leq n}, \;\; \mbox{ for each }
n \geq 1.
$$
\end{theorem}
\proof
Let $p_i$, $i = 1, 2, \ldots$, be the $i$th prime.
We define
an $n$-state ($n \geq 1$) transducer
$T_n = (\{1, \ldots, n \}, 1, \Delta_n)$ by setting
by $\Delta_n(1, 0) = (1, 0^{p_n})$, $\Delta_n(i, 0) = (i, \varepsilon)$,
$2 \leq i \leq n$, $\Delta_n(j, 1) = (j+1, \varepsilon)$,
$1 \leq j \leq n-1$, and $\Delta_n(n, 1) = (n, \varepsilon)$.

In the encoding $S_1$ we use the string $\sigma_n = {\rm bin}(n)$ to encode
the transducer $T_n$, $n \geq 1$.
Any transducer $T$ that is not one of the
above transducers $T_n$, $n \geq 1$, is encoded in
$S_1$ by a string $0 \cdot e$, $e \in \{ 0, 1 \}^*$, where $|e|$
is at least the  sum of the lengths of outputs
of all transitions in $T$. This condition is satisfied, for example
by choosing the encoding of $T$ in $S_1$ to be simply 0 concatenated
with the
standard encoding of $T$.

Let $m \geq 1$ be arbitrary but fixed. The string $0^{p_m}$ has a description
$(T^{S_1}_{\sigma_m}, 0)$ of size $\lceil \log m \rceil + 1$,
where $\sigma_m \in S_1$ encodes $T_m$ and
the transducer $T^{S_1}_{\sigma_m}$ has $m$ states.
We show that $C_{S_1}(0^{p_m}) = \lceil \log m \rceil + 1$.

By the definition of the transducers $T_n$, for any $w \in \{ 0, 1 \}^*$,
$T_n(w)$ is of the form $0^{k \cdot p_n}$, $k \geq 0$. Thus,
$0^{p_m}$ cannot be the output of any transducer $T_n$, $n \neq m$.

On the other hand, consider an
arbitrary description $(T^{S_1}_\sigma, w)$
of the string $0^{p_m}$ where $T^{S_1}_\sigma$ is not any
of the transducers $T_n$, $n \geq 1$. Let $x$ be the length
of the longest output of a transition of $T^{S_1}_\sigma$.
Thus, $x \cdot |w| \geq p_m$. By the definition of $S_1$ we
know that $|\sigma| \geq x + 1$, and we conclude that
$$
|| (T^{S_1}_\sigma, w) ||_{S_1} = |\sigma| + |w| > \lceil \log m \rceil + 1.
$$
We have shown that, in the encoding $S_1$,
the unique minimal description
of $0^{p_m}$ uses a transducer with $m$ states, which
implies $0^{p_m} \in L^{S_1}_{ =m}$.
\endpf

The encoding $S_1$ constructed in the proof of Theorem~\ref{se-tatta2}
is not a polynomial-time encoding because $T_n$ has an encoding
of length $O(\log n)$, whereas the description of the transition
function of $T_n$ (in the format specified in Section~\ref{prelim})
has length $\Omega(n \cdot \log n)$. Besides  the above
problem $S_1$ is otherwise efficiently computable and using  standard
``padding techniques'' we can simply increase the length of all
encodings of transducers in $S_1$.

\begin{corollary}
\label{se-tatta3}
There exists a polynomial time encoding
$S_{1}'$
such that
$$
L^{S_{1}'}_{ \leq n-1} \subset L^{S_{1}'}_{ \leq n}, \;\; \mbox{ for each }
n \geq 1.
$$
\end{corollary}

\proof
The encoding $S_1'$ is obtained by modifying the encoding $S_1$ of
the proof of Theorem~\ref{se-tatta2} as follows. For $n \geq 1$,
$T_n$ is encoded by the string $\sigma_n =
{\rm bin}(n)^{\dagger} \cdot 1^n$.
Any transducer $T$ that is not one of the transducers $T_n$, $n \geq 1$,
is encoded by a string $0 \cdot w$ where $|w| \geq 2^x$ and $x$
is the sum of the lengths of outputs of all transitions of $T$. If
$\sigma$ is the standard encoding of $T$, for example, we can choose
$w = \sigma^{\dagger} \cdot 1^{2^{|\sigma|}}$.

Now $|\sigma_n|$ is polynomially related to the length of the description
of the transition function of $T_n$, $n \geq 1$, and given
$\sigma_n$ the transition function of $T_n$ can be output in quadratic time.
For transducers not of the form $T_n$, $n \geq 1$, the same holds trivially.

Essentially in the same way as in the proof of Theorem~\ref{se-tatta2},
we verify that for any $m \geq 1$, 
the string $0^{p_m}$ has a unique minimal
description $(T^{S_1'}_{\sigma_m'}, 0)$, where
$\sigma_m' \in S_1'$ is the description of the $m$-state
transducer $T_m$. The same argument works because, the encoding
of any transducer $T$ in $S_1'$ is, roughly speaking, obtained from
the encoding $\sigma$ of $T$ in $S_1$ by appending $2^{|\sigma|}$
symbols 1.
\endpf

There exist computable encodings that allow minimal descriptions of
strings based on transducers with different numbers of states.
Furthermore, the gap between the numbers of states of
the transducers used for different minimal descriptions
of the same string can be made arbitrarily
large, that is,
for any $n < m$ we can construct an encoding where
some string has  minimal descriptions both using
transducers with either $n$ or $m$ states. 
The proof  uses an idea
similar to the proof of Theorem~\ref{se-tatta2}.

\begin{theorem}
\label{se-tatta4}
For any $1 \leq n < m$, there exists a computable encoding
$S_{n, m}$ such that $L^{S_{n,m}}_{\exists_{\rm min} m} \cap L^{S_{n, m}}_{=n} 
\neq \emptyset$.
\end{theorem}

Note that the statement of Theorem~\ref{se-tatta4} implies that
$L^{S_{n, m}}_{=m} \neq L^{S_{n,m}}_{\exists_{\rm min} m}$.
Again, by padding the encodings as in Corollary~\ref{se-tatta3},
the result of Theorem~\ref{se-tatta4} can be established using
a polynomial-time encoding.

\section{Conclusion}

As perhaps expected, the properties of the state-size hierarchy
with respect to the specific  computable encodings considered
in section~\ref{sec-viisi} could be established
using constructions where we  added to transducers additional states
without changing the size of the encoding. 
In a similar way
various other properties can be established
for the state-size hierarchy corresponding 
to specific (artificially defined)
computable encodings. The main open problem concerning general
computable encodings is whether it is possible
to construct an encoding
for which the state-size hierarchy collapses
to some finite level, see Problem~\ref{se-tatta}.

As our main result we have established that the state-size hierarchy
with respect to the standard encoding is infinite.
Many interesting open problems dealing with the hierarchy
with respect to the standard encoding remain. In addition to the problems
discussed in section~\ref{statessize}, we can consider various types
of questions related to combinatorics on words. For example,
assuming that a minimal description of a string $w$ needs a transducer
with at least $m$ states, is it possible that $w^2$ has a minimal
description based on a transducer with less than $m$ states?

\begin{conjecture}
If $w \in L_{=m}$ ($m \geq 1$), then for any $k \geq 1$,
$w^k \not\in L_{\leq m-1}$.
\end{conjecture}


\end{document}